\documentclass[a4paper,12pt]{article}

\usepackage{amsmath,amssymb,amsfonts} 
\usepackage{graphicx}                 
\usepackage{subcaption}
\usepackage{float}
\usepackage[space]{grffile}
\usepackage{color}                    
\usepackage{hyperref}                 
\definecolor{red}{rgb}{1,0,0}           
\definecolor{green}{rgb}{0,1,0}
\definecolor{blue}{rgb}{0,0,1}
\definecolor{darkblue}{rgb}{0,0,0.5}
\definecolor{lightblue}{rgb}{.5,.5,1}
\definecolor{lightgray}{gray}{.87}          
\definecolor{Dark}{gray}{.20}
\definecolor{pink}{rgb}{.95,0.82,0.92}  
\definecolor{yellow}{rgb}{1,1,0}
\definecolor{lightyellow}{rgb}{1,1,.5}
\definecolor{purple}{rgb}{0.7,0,0.85}
\definecolor{darkgreen}{rgb}{0,0.5,0}
\definecolor{orange}{rgb}{0.8,0.2,0.2}
\def \be {\begin{equation}}
\def \ee {\end{equation}}
\def \bea {\begin{eqnarray}}
\def \eea {\end{eqnarray}}
\def \nn {\nonumber}

\def \rr {\raise.35ex\hbox{\small $\prime$}\kern-.17em{\mbox{\large $\imath$}}}
\def \del {\partial}
\def \dels {\partial\kern-.5em / \kern.5em}
\def \As {{A\kern-.5em / \kern.5em}}
\def \Ds {D\kern-.7em / \kern.5em}

\newcommand{\detail}[1]{}

\newcommand{\hide}[1]{}

\pdfoutput=1

\begin{document}

\pagestyle{plain}


\begin{titlepage}

\begin{center}

\noindent
\textbf{\LARGE
Asymptotic States of Black Holes
\\
\vskip 1em
in KMY Model
}
\vskip 4em
{\large 
Pei-Ming Ho,${}^{a}$
\footnote{e-mail address: pmho@phys.ntu.edu.tw}
Yoshinori Matsuo,${}^{a, b}$
\footnote{e-mail address: matsuo@het.phys.sci.osaka-u.ac.jp}
Shu-Jung Yang${}^{a}$
\footnote{e-mail address: r05222083@ntu.edu.tw}
}
\\
{\vskip 10mm \sl

${}^{a}$
Department of Physics and Center for Theoretical Sciences, \\
Center for Advanced Study in Theoretical Sciences, \\
National Taiwan University, Taipei 106, Taiwan,
R.O.C. 

${}^{b}$Department of Physics, Osaka University, Toyonaka, Osaka 560-0043, Japan
}\\

\vskip 3mm
\vspace{60pt}
\begin{abstract}

Following the work of Kawai, Matsuo, and Yokokura,
we study the dynamical collapsing process with spherical symmetry
in the time-dependent space-time background
including the back-reaction of Hawking radiation.
We show that in this model there are two classes of asymptotic solutions.
One of the two classes is known previously.
These states have the slope $\del a/\del r$ approximately equal to $1$.
The other class of asymptotic solutions
is that of shells with a small thickness.
We emphasize that these thin shells should be properly
understood as configurations in the low-energy effective theory.
They behave characteristically differently
from the singular states of ideal thin shells of zero thickness.

\end{abstract}
\end{center}
\end{titlepage}

\setcounter{page}{1}
\setcounter{footnote}{0}
\setcounter{section}{0}


\section{Introduction}

In the study of the black holes,
it is common practice to study the formation and evaporation of a black hole
separately as independent processes
for the simplicity of calculation.
During the formation process,
which is typically treated as a purely classical process,
classical matter collapses and an event horizon appears.
After that,
the evaporation process due to the quantum effect 
is considered as an independent process.
In this approximation scheme,
the evaporation is computed in the presence
of the event horizon of a classical black hole
with a constant Schwarzschild radius.

Of course,
in reality,
the Schwarzschild radius must decrease over time
if the black hole completely evaporates in the end.
But some people argued that the extremely slow change 
in the Schwarzschild radius can be ignored
for the study of the black-hole evaporation.
As a test of the robustness of these arguments,
one should check whether the evaporation
is significantly modified if the time-dependence
of the Schwarzschild radius is turned on.

This was done in the paper of Kawai, Matsuo and Yokokura \cite{Kawai:2013mda}.
(See also Refs.\cite{Kawai:2014afa} | \cite{Kawai:2017txu}.)
The formation and evaporation of a black hole 
is viewed as a single process,
and the back-reaction of Hawking radiation on the geometry is included
from the very beginning of the black-hole formation.
The Schwarzschild radius is time-dependent
due to Hawking radiation.

It turns out that,
as the vacuum energy-momentum tensor is assumed to be 
dominated by the Hawking radiation
outside the collapsing matter
in this model,
when the collapsing matter has a smooth density distribution,
it completely evaporates without apparent horizon \cite{Kawai:2013mda}.
In contrast,
an ideal thin shell with a delta-function energy density
does not evaporate completely.
Instead,
it approaches a classical black hole in the infinite future
with an event horizon \cite{Kawai:2013mda}.

There are a few obvious questions.
For example,
is the ideal thin shell of delta-function energy distribution physical?
Are there other classes of behaviors characteristically different
from the two classes of solutions already found in Ref.\cite{Kawai:2013mda}?
These are the questions that motivated this research project.

In this paper,
we first discuss the difference between 
the thin shell with delta function energy distribution 
and that with a finite but very small thickness. 
These two shells are expected to be almost indistinguishable.
However, they have characteristically different behaviors in calculations of the Hawking radiation. 
We shall demonstrate the difference by comparing a thin shell 
with a thickness of the Planck scale
and a shell with a delta function energy distribution. 
While the latter survives in the end as a classical black hole,
the former evaporates completely within finite time.
Thus, the thin shell with delta function density distribution 
is unphysical, and hence we will focus on 
the configurations with a finite thickness no less than the Planck scale. 

Then, 
we will show that there are at least two classes of asymptotic behaviors
of the collapsing matter 
as a result of the back-reaction of Hawking radiation.
\footnote{
Recall that 2D black holes are also categorized as two classes.
It was shown in Ref.\cite{Russo:1992ax}
that an apparent horizon will either appear 
or be absent depending on the magnitude of the ingoing energy flux.
}
One of the two classes has already been proposed and studied in
Ref.\cite{Kawai:2013mda} and Refs.\cite{Kawai:2014afa} | \cite{Kawai:2017txu}.
The other class is described as a thin shell in the low-energy theory,
and it should be distinguished from the ideal thin shell of zero thickness.
We will examine the details of both classes of collapsing processes in this paper
through both analytical study and numerical simulation.

The plan of this paper is as follows. 
In Sec.\ref{Review-KMY}, 
we briefly review the KMY model.
In Sec.\ref{ThinShellSec},
we examine the notion of thin shells
in the context of low-energy effective theories.
We argue that it is inappropriate to apply 
the low-energy theory to ideal thin shells of zero thickness,
as their behavior is characteristically different from 
thin shells of a small but finite thickness.
In Sec.\ref{SmoothConfig},
we consider shells with an energy distribution 
for which the slope $\del a/\del r \simeq 1$.
These configurations are particularly interesting
as asymptotic states \cite{Kawai:2013mda}.
Excluding the unphysical ideal thin shells,
in Sec.\ref{AS},
we argue that there are two types of asymptotic states,
the slope-1 states and the thin shell states.
This claim is backed up by numerical simulation
presented in Sec.\ref{Numerical}.
Finally, 
we conclude in Sec.\ref{Conclusion}.


\section{Review of KMY Model}
\label{Review-KMY}

We will refer to the approach of Ref.\cite{Kawai:2013mda},
which is followed by Refs.\cite{Kawai:2014afa}--\cite{Kawai:2017txu},
as the KMY model.
One of the crucial points of this approach is that
the back-reaction of Hawking radiation to
the geometry is taken into account.
But the importance of the back-reaction of Hawking radiation,
or that of the vacuum energy-momentum tensor,
has been considered before the KMY model
\cite{HR1}--\cite{Paranjape:2009ib} 
in the literature.
(See also Refs.\cite{Mersini-Houghton:2014zka}--\cite{Bardeen:2014uaa}
for later proposals.)
Another crucial feature of the KMY model is that
it further assumes that the vacuum energy-momentum tensor
is dominated by Hawking radiation.

Unlike most of the models of black holes,
the null energy condition is not violated in the KMY model,
and this is directly related to the absence of the apparent horizon.
(See e.g. Ref.\cite{Ho:2019kte}.)
Strictly speaking,
the violation of the null energy condition 
in conventional models of black holes
is based on a few assumptions.
For instance,
it is often considered as a consequence of the equivalence principle.
However,
in general, the quantum states cannot be defined locally.
Instead, they depend on the boundary conditions. 
The quantum state of the black hole 
can break the local equivalence principle,
and the vacuum energy-momentum tensor should be determined
by a specific quantum field theory which has a consistent UV limit
including quantum gravity.
It is not clear whether certain important quantum gravity effects
are missing from the quantum field theories people usually consider
in their models.
Furthermore,
as it was pointed out in Ref.\cite{Mathur:2009hf},
there must be ``drama'' at the horizon for the information to be preserved,
as it was also argued in Ref.\cite{Almheiri:2012rt}.

Our viewpoint is that the quantum state breaks the equivalence principle
to admit unitarity for a UV-complete theory.
The vacuum energy-momentum tensor obviously depends on the matter content of the model. 
While the vacuum energy-momentum tensor is typically calculated 
in some simplified models of matter fields 
(e.g. 2 dimensional massless scalar fields of s-wave approximation in Ref.\cite{Davies:1976ei})
in the conventional model, 
the KMY model simply assumes the vacuum energy-momentum tensor
to be dominated by the outgoing positive energy flux
(i.e. the Hawking radiation)
as an alternative.
It turns out that,
as a consequence of the semi-classical Einstein equation,
a pressure at the Planck scale \cite{Kawai:2014afa}
appears on the collapsing shell as an effective ``firewall''.
This is consistent with the analysis of Ref.\cite{Arderucio-Costa:2017etb},
which shows that a pressure-less thin shell is inconsistent
with the absence of horizon,
and in agreement with Ref.\cite{Mann:2018jcf},
which says that either pressure or charges are necessary to keep the shell null.

While there is no good reason to strictly preserve
the null energy condition in a quantum field theory for all quantum states,
it may be only weakly violated to the extent
that the KMY model is still a good approximation.
(See Ref.\cite{Kawai:2017txu} for a generalization of the KMY model
with a more general vacuum energy-momentum tensor.)
To say the least, 
the KMY model is an interesting alternative to
conventional models of black holes that
may provide a self-consistent story including quantum effects.

In this paper,
we are interested in the asymptotic configurations
in the KMY model.
It was shown in Ref.\cite{Kawai:2013mda} that
there would be no event or apparent horizon
\footnote{
The possibility that black holes have no horizon
has also been proposed by many others.
For an incomplete list, see Refs.
\cite{Gerlach:1976ji} | \cite{FuzzBall},
\cite{Banerjee:2002rv}, \cite{Vachaspati:2006ki},
\cite{Barcelo:2007yk} | \cite{Fayos:2011zza},
\cite{Mersini-Houghton:2014zka},
\cite{Mersini-Houghton:2014cta},
\cite{Saini:2015dea}, \cite{Baccetti:2017oas}.
}
for a certain smooth configuration,
which will be referred to as the ``slope-1'' configuration in this paper.
\footnote{
It was called an ``asymptotic black hole'' in Ref.\cite{Ho:2016acf}.
}
In fact,
it can be proven \cite{Ho:2015fja,Ho:2015vga}
that there is no apparent horizon as long as 
the collapsing matter completely evaporates within a finite time.
According to the semiclassical Einstein equation,
the Schwarzschild radius shrinks with time 
in a superluminal fashion
due to the loss of energy into Hawking radiation.
As a result,
the collapsing matter can never
fall through the Schwarzschild radius,
as long as the (incipient) black hole
evaporates completely within finite time \cite{Kawai:2013mda,Ho:2015fja}.

An exceptional configuration for the collapsing matter 
that does not evaporate within a finite time
is the case of the thin shell 
with an energy density given by the Dirac delta function.
Its Hawking radiation decreases with time so that
the black hole survives in the infinite future,
and an event horizon arises like a classical black hole
\cite{Kawai:2013mda}.

Notice that the time evolution of a thin shell 
of absolutely zero thickness
may or may not be obtained by the zero thickness limit from 
a thin shell of finite thickness,
as the equation determining Hawking radiation involves higher derivatives. 
We investigate in this paper whether
small (Planck-scale) deviations from the mathematical notion of
a perfect zero-thickness thin shell
would lead to characteristically different 
space-time structures at large scales.

Another specific question we would like to answer is 
whether the smooth configuration 
studied in Ref.\cite{Kawai:2013mda}
(called the ``slope-1'' configuration)
is the only asymptotic limit for generic initial states.
We shall find that there is another class of asymptotic configurations.

\subsection{Metric}

A generic, spherically symmetric 4D metric can be put in the 
following form 
in the outgoing Eddington-Finckelstein coordinate 
\be
ds^2 = - e^{2\psi(u, r)}\left(1-\frac{a(u, r)}{r}\right) du^2
- 2 e^{\psi(u, r)} du dr + r^2 d\Omega^2,
\label{metric}
\ee
which involves two parametric functions $\psi(u, r)$ and $a(u, r)$. 
The time evolution of this metric will be studied in terms of 
the Eddington retarded time $u$.
Due to spherical symmetry,
the functions $\psi(u, r)$ and $a(u, r)$ only depend
on $u$ and the areal radius $r$.

At any given instant of time $u$,  
the function $a(u, r)$ in the metric \eqref{metric}
gives twice the Bondi mass
inside the ball of the radius $r$ centered at the origin.
The other function $\psi(u, r)$ in the metric \eqref{metric}
is the exponent of the redshift factor $e^{\psi(u, r)}$
between the retarded time coordinate $u$ at spatial infinity
and the retarded time coordinate $\hat{u}(u, r)$ at $r$ \cite{Ho:2016acf}.

The metric \eqref{metric} is only suitable for the region $r > r^{\ast}(u)$,
where $r^{\ast}(u)$ is the (largest) solution to the equation $r^{\ast}(u) = a(u, r^{\ast}(u))$.
It is also only valid outside the apparent horizon.
As the apparent horizon appears in most models of black holes,
the metric \eqref{metric} is often considered inappropriate,
and so the coordinate system using the $(v, r)$ coordinates
(with the ingoing Eddington-Finckelstein coordinate)
is more commonly used in the literature.
However,
as we mentioned in the introduction,
there is no apparent horizon in the KMY model,
as a consequence of the non-violation of the null energy condition,
as opposed to the conventional model.
(It has been shown in Refs.\cite{Baccetti:2019mab,Ho:2019kte}
that the null energy condition has to be violated for the existence of the apparent horizon.)
In any case,
one can use any metric until a (coordinate) singularity appears.
The metric \eqref{metric} will turn out to be convenient
in the discussion below for the KMY model.

We assume that the collapsing 
matter
has an outer radius $R_0(u)$
beyond which there is no ingoing energy flux
(but there can be outgoing energy flux as Hawking radiation),
so that $a(u, r)$ is $r$-independent outside the outer radius $R_0(u)$:
\be
a(u, r) = a_0(u) \quad \mbox{for} \quad r \geq R_0(u)
\label{a=a0}
\ee
for some function $a_0(u)$
which is twice the total Bondi mass of the collapsing matter.

When $a_0(u)$ is time-independent,
as in the classical case without Hawking radiation,
the metric \eqref{metric} is equivalent to 
the Schwarzschild metric with Schwarzschild radius $a_0$
for $r > R_0(u)$.
When $a_0(u)$ is not time-independent,
there is outgoing energy flux for $r > R_0(u)$ given by
\be
T_{uu} = - \frac{1}{\kappa r^2} \frac{da_0(u)}{du},
\label{Tuu=da0du}
\ee
where $\kappa = 8\pi G_N$ and $G_N$ is the Newton constant.
This outgoing energy flux is used to represent Hawking radiation
and is assumed to be positive,
so that the Schwarzschild radius $a_0(u)$ decreases over time
and $da_0/du < 0$.

For simplicity,
in this paper, we shall assume that 
the collapsing 
matter
is falling at the speed of light.
Applying the general results of Ref.\cite{Ho:2016acf} to this special case,
the redshift factor $e^{\psi(u, r)}$ is given by
\be
\psi(u, r) =
- \int_{r}^{\infty} d\bar{r} \;
\frac{\frac{\del a(u, \bar{r})}{\del \bar{r}}}{\bar{r} - a(u, \bar{r})}.
\label{psiur}
\ee
Because of eq.\eqref{a=a0},
$\psi(u, r) = 0$ for $r > R_0(u)$.

According to eq.\eqref{psiur},
the redshift factor for the retarded time $U$ of the Minkowski space
inside the collapsing 
matter is
\be
\log\left(\frac{dU(u)}{du}\right)
= \psi_0(u) =
- \int_{0}^{R_0(u)} dr \; \frac{\frac{\del a(u, r)}{\del r}}{r - a(u, r)}.
\label{psi0}
\ee
Since
$a(u,r)$ is always a monotonic function of $r$
in the range of integration in eq.\eqref{psi0}, 
$a$ can be 
used as a coordinate in place of $r$ to parametrize
the integral,
hence
the integral \eqref{psi0} can also be expressed as 
\be
\psi_0(u) = - \int_{0}^{a_0(u)} \frac{da}{r(u,a) - a}. 
\label{psi0a}
\ee

One may wonder if $\psi_0(u)$ diverges in the limit $r\rightarrow a$.
It is impossible as long as there is no divergence in the Hawking radiation, 
as we will discuss later in the paragraph below eq.\eqref{19}.
This also means that $r$ can never coincide with $a$,
and thus there would be no apparent horizon.

\subsection{Hawking Radiation}

The Hawking radiation is created during the gravitational collapse
because the quantum vacuum state of 
incoming matter in the infinite past
evolves to a state
that is no longer the vacuum state at large $r$
after the gravitational collapse. 
We shall adopt the formula for Hawking radiation of Ref.\cite{Kawai:2013mda},
which is in agreement with that of Refs.\cite{Davies:1976hi,Davies:1976ei},
although the rest of the vacuum energy-momentum tensor is omitted.

Following these works, 
approximating the vacuum energy-momentum tensor by 
that of s-wave modes of massless scalar fields, and 
assuming that the initial state in the infinite past
is the Minkowskian vacuum state,
the Hawking radiation is given by the energy flux
\be
T_{uu} = \frac{{\cal N}}{4\pi r^2} \{u, U(u)\},
\label{Tuu}
\ee
where $\kappa = 8\pi G$,
${\cal N}$ is a numerical constant proportional to 
the number of massless fields in Hawking radiation
and $\{u, U(u)\}$ is the Schwarzian derivative defined by
\be
\{u, U(u)\} \equiv
\left[\frac{\frac{d^2 U(u)}{du^2}}{\frac{dU(u)}{du}}\right]^2 
- \frac{2}{3}\frac{\frac{d^3 U(u)}{du^3}}{\frac{dU(u)}{du}}.
\ee

Using eq.\eqref{psi0},
one can rewrite the Schwarzian derivative as
\be
\{u, U(u)\} \equiv
\frac{1}{3}\left[\dot{\psi}_0^2(u) - 2\ddot{\psi}_0(u)\right],
\label{uU-Schwarzian}
\ee
where the dots denote $u$-derivatives.

\section{Thin Shells}
\label{ThinShellSec}

In this section, 
we first review the case of an ideal thin shell,
i.e. a thin shell of zero thickness.
Then we compare it with the notion of a pseudo thin shell,
which has 
a finite thickness larger than
the cutoff length scale of the low-energy effective theory,
but it is thin enough so that
its behavior is not sensitive to the precise thickness.
Their behaviors turn out to be characteristically different.

\subsection{Ideal Thin Shell}
\label{ThinShell}

For a thin shell of zero thickness,
the function $a(u, r)$ in the metric \eqref{metric} is given by
\be
a(u, r) = a_0(u)\Theta(r-R_0(u)),
\label{a-Theta}
\ee
where $\Theta(x)$ is the step function that is $0$ or $1$
for $x < 0$ or $x > 0$.
It has a diverging energy density 
proportional to the Dirac delta function
that diverges at $r = R_0(u)$.
For the low-energy effective theory to be applicable,
a shell should have a finite thickness
much larger than the Planck length,
and an energy density much smaller than the Planck scale.
However,
this notion of an ideal thin shell has been 
widely used in the literature
in the context of low-energy effective theories.

To evaluate $\psi_0(u)$,
we use eq.\eqref{psi0a},
with $r(u, a)$ given by inverting eq.\eqref{a-Theta}:
\begin{equation}
 r(u,a) = R_0(u) \ . 
\end{equation}
Hence eq.\eqref{psi0a} can be evaluated as
\be
\psi_0(u) = - \int_{0}^{a_0(u)} \frac{da}{R_0(u) - a}
= \log\left(\frac{R_0(u)-a_0(u)}{R_0(u)}\right).
\label{psi0-ts}
\ee

\subsubsection{Background With Constant Schwarzschild Radius}

Let us first consider the evolution of the ideal thin shell
with a fixed Schwarzschild radius,
ignoring the back-reaction of Hawking radiation.

The thin shell is by assumption falling at the speed of light,
hence we have
\be
\frac{dR_0(u)}{du} = - \frac{1}{2}\frac{R_0(u) - a_0(u)}{R_0(u)}
\label{dR0du}
\ee
according to the metric \eqref{metric}.
When $a_0$ is assumed to be time-independent,
its solution is
\be
R_0(u) \simeq a_0 + C_0 e^{-\frac{u}{2a_0}}
\label{R0-sol}
\ee
when $R_0(u) - a_0 \ll a_0$.
Eq.\eqref{psi0-ts} then gives
\be
\dot{\psi}_0(u) \simeq - \frac{1}{2a_0},
\label{dotpsi0}
\ee
and the Schwarzian derivative can be easily computed
\be
\{u, U(u)\} = \frac{1}{3}\left(
\dot{\psi}_0^2 - 2 \ddot{\psi}_0
\right)
\simeq \frac{1}{12a_0^2},
\label{12a02}
\ee
which gives the conventional result of Hawking radiation
\be
\frac{da_0}{du} \simeq - \frac{\kappa{\cal N}}{48\pi a_0^2}.
\label{da0du-1}
\ee
As a result,
the thin shell evaporates completely within a finite time
of order $\mathcal{O}(a_0^3/\kappa)$.

Some people have argued that a constant background is justified as 
a good approximation because the time scale of the change in
the Schwarzschild radius is $\mathcal{O}(a_0^3/\kappa)$,
while the time-scale of the gravitational collapse is $\mathcal{O}(a_0)$;
the large hierarchy in the time scales implies that the former
cannot have a significant effect on the latter.
This argument is not rigorous because,
strictly speaking,
it makes sense to compare the time scales only when
both are defined with respect to the same observers.
The time scale of the Schwarzschild radius is $\mathcal{O}(a_0^3/\kappa)$
for distant observers,
and yet the time-scale of gravitational collapse is $\mathcal{O}(a_0)$
only for infalling observers.
Classically,
it takes an infinite time for distant observers to see
the star falling inside the horizon,
so after including the quantum effect,
the time scale of gravitational collapse should also be
$\mathcal{O}(a_0^3/\kappa)$ for distant observers.
Indeed,
we will see below that 
the time-dependence of the Schwarzschild radius,
despite of how small it is,
can have a significant effect.

\subsubsection{Back-Reacted Background}
\label{idealthinshell-BR}

However,
as there is Hawking radiation,
$a_0(u)$ must decrease over time due to 
the outgoing energy flux 
\eqref{Tuu=da0du}. 
Since the Hawking radiation is given by \eqref{Tuu}, 
the time evolution of $a_0(u)$ is given by the differential equation; 
\be
\frac{da_0(u)}{du} = - \frac{\kappa {\cal N}}{4\pi} \{u, U(u)\}.
\label{da0du}
\ee
This modification would change the evaluation of $\dot{\psi}_0(u)$ \eqref{dotpsi0}.
Using eqs.\eqref{psi0-ts} and \eqref{dR0du},
we find
\be
\dot{\psi}_0(u)
\simeq - \frac{a_0(u)}{2R_0^2(u)} - \frac{\dot{a}_0(u)}{R_0(u)-a_0(u)}.
\label{19}
\ee
While the first term is always finite,
the 2nd term diverges at the horizon
unless $\dot{a}_0(u)$ vanishes at the horizon.

As the outgoing energy flux $T_{uu}$ of the energy-momentum tensor
is identified with the Hawking radiation in the KMY model,
the finiteness of $T_{uu}$ implies,
through eqs.\eqref{Tuu} and \eqref{uU-Schwarzian},
that $\dot{\psi}_0$ must be finite,
unless its divergence cancels the divergence in the other term in eq.\eqref{uU-Schwarzian}.
The cancellation of divergence in eq.\eqref{uU-Schwarzian} demands
$\dot{\psi}_0 \sim \frac{2}{c - u}$ for some constant $c$ as $u \rightarrow c$,
with the implication that $\dot{\psi}_0$ is positive when $u\rightarrow c$ (but $u < c$).
However,
$\dot{\psi}_0$ should be negative according to eq.\eqref{psi0},
since the factor $dU/du$ should be decreasing with time.
Hence it is impossible for $\dot{\psi}_0$ to blow up.
Thus, 
it is inconsistent to have a diverging $\dot{\psi}_0$ at the horizon,
hence $\dot{a}_0$ must vanish at the horizon.
The conclusion is thus that
once the time-dependence of the Schwarzschild radius $a(u)$ is taken into account,
the ideal thin shell cannot cross the horizon 
without turning off Hawking radiation.
Indeed,
it was shown in Ref.\cite{Kawai:2013mda} that
the Hawking radiation decreases to zero 
and a classical black hole survives the incomplete evaporation
for a collapsing ideal thin shell.

More explicitly, 
assuming that $a_0(u)$ changes very slowly, 
eq.\eqref{dR0du} implies that the shell 
asymptotes to the Schwarzschild radius $r=a_0(u)$. 
When the shell is sufficiently close to the Schwarzschild radius 
$R_0(u) \simeq a_0(u)$,
eq.\eqref{da0du} is solved with \cite{Kawai:2013mda}
\begin{align}
u &\simeq
\frac{e^{-\frac{D^2}{2}}}{6\pi B} \int_D^{\xi} d\xi' \; e^{\frac{{\xi'}^2}{4}},
\\
a(u) &\simeq
a(0) - B \int_D^{\xi} d\xi' \; e^{-\frac{{\xi'}^2}{4}},
\end{align}
for constant parameters $B$ and $D$.
This solution describes a decaying Hawking radiation 
that vanishes at the event horizon,
and the ideal thin shell is only partially evaporated.
See Ref.\cite{Kawai:2013mda}
also for numerical simulation.

To be more precise,
the collapsing shell exponentially approaches the shell
as described by eq.\eqref{R0-sol} 
for constant Schwarzschild radius
when the back-reaction from Hawking radiation is ignored. 
In the back-reacted geometry,
eq.\eqref{R0-sol} is modified as 
\be
R_0(u) \simeq a_0(u) + C_0 e^{-\frac{u}{2a_0(u)}} - 2 a_0(u) \dot a_0(u) \ . 
\ee
As $\dot{a}_0(u) \leq 0$,
the shell cannot reach the Schwarzschild radius unless
$\dot{a}_0(u) \rightarrow 0$ \cite{Kawai:2013mda}. 
By using the conventional formula of the Hawking radiation, 
$\dot a_0 = - \sigma/a_0^2$,
the minimum distance after a long time 
would be 
\be
R_0(u) \simeq a_0(u) + \frac{2 \sigma}{a_0(u)} \ . 
\ee
This argument is generalized in Ref.\cite{Baccetti:2017oas}
to the general metric \eqref{metric}
with finite and continuous $\psi(u,r)$
and with positive definite $a(u,r)$ before the complete evaporation.
(See the reference for more detailed conditions on the geometry.)

To summarize,
for the ideal thin shell,
if we ignore the time-dependence of the Schwarzschild radius,
the black hole would completely evaporate as in the conventional model,
but if we account for the time dependence of the Schwarzschild radius,
the black hole would not evaporate completely.
While one may suspect that certain omitted details of the quantum effect 
involved in this process could further change the conclusion,
it is,
to say the least,
an example showing that
the back-reaction of Hawking radiation has the potential to play a crucial role.
Calculations without back-reaction need to be further justified.


\subsection{Pseudo Thin Shell}
\label{ThinShellPlanck}

Here we discuss the notion of a ``pseudo thin shell''
in the context of the low-energy effective theory
with a cutoff length scale $\ell$
larger than the Planck length $\ell_p$. 
It turns out that the behavior of the pseudo thin shell 
is characteristically different from the ideal thin shell
(when the time-dependence of the background is turned on).
This means that we cannot trust the low-energy effective theory
on its description of the ideal thin shell.
In the context of
low-energy effective theories,
the notion of the ideal thin shell should be viewed as invalid.

The purpose of this subsection is to point out
the fact that the ideal thin shell
is over-sensitive to details at the Planck scale.
For this purpose,
we do not have to justify our choice of the profile
for the pseudo thin shell.
Nevertheless,
the pseudo thin shell is a natural consideration
as an interpolation between
the ideal thin shell and the slope-1 configuration described
in the next section.

It was pointed out in Ref.\cite{Kawai:2013mda} that
the Hawking radiation from a single ideal thin shell
and that from the continuum limit of infinitely many shells
(the slope-1 configuration) are very different. 
This may seem weird at first sight 
since the slope-1 configuration was constructed as a collection of many thin shells.
The reason behind is that the ideal thin shell
has pathological behavior due to higher-derivative terms in its evolution equation,
which can be removed in a suitable continuum limit.

When we model a smooth configuration as
a large (but finite) number of thin shells in numerical simulation,
it is better to consider a collection of pseudo thin shells
(instead of ideal thin shells)
to avoid the pathological behaviors of ideal thin shells.
In terms of pseudo thin shells,
that there is no drastic difference between
the Hawking radiation for
a single (pseudo thin) shell and 
that for a configuration of many thin shells
(e.g. the slope-1 configuration). 
We will have more discussions on the pseudo thin shell
in the next subsection.

Here, we consider the pseudo thin shell with the following profile; 
\be
a(u, r) \simeq
\left(r - \frac{2\sigma}{r}\right)\left[\Theta(r-R_i(u))-\Theta(r-R_0(u))\right]
+ a_0(u)\Theta(r-R_0(u)),
\label{profile-3}
\ee
where $R_0$ and $R_i$ are the outer and inner radii,
and $R_0(u) - R_i(u)$ is assumed to be much smaller than $R_0(u)$.
This pseudo thin shell still contains an ideal thin shell 
at the innermost surface. 
While this does not interfere with our purpose to show
that a small deviation from an ideal thin shell makes a large difference,
the difference between an ideal thin shell only at the innermost surface
and a finite-density distribution everywhere is 
negligible since physics at the innermost part is 
almost irrelevant due to the large redshift factor as we will see soon. 

The Schwarzschild radii at the outer and inner surfaces of the shell are thus
\begin{align}
a_0(u) &\equiv R_0(u) - \frac{2\sigma}{R_0(u)},
\\
a_i(u) &\equiv R_i(u) - \frac{2\sigma}{R_i(u)}.
\end{align}
These formulas, together with 
the assumption that the shells are collapsing at the speed of light,
imply that 
the Schwarzschild radii $a(u)$ change with time 
according to the conventional formula of Hawking radiation
$\dot{a} \simeq - \sigma/a^2$ 
for some constant $\sigma$.
In comparison with eq.\eqref{da0du-1},
$\sigma$ is a constant parameter of the Planck scale
\be
\sigma \simeq \frac{\kappa{\cal N}}{48\pi} \sim \mathcal{O}(\ell_P^2),
\label{sigma}
\ee
where ${\cal N}$ was defined in eq.\eqref{Tuu}.
We shall compare the rates of evaporation
for a pseudo thin shell and an ideal thin shell,
taking into consideration
the time-dependence of the Schwarzschild radius.

Within the shell ($r \in (R_i(u), R_0(u)]$ ),
except for the innermost surface of the shell, 
this is the same profile as the smooth configuration 
proposed in Ref.\cite{Kawai:2013mda},
which will be discussed below in Sec.\ref{SmoothConfig}.
But at the same time, 
at a length scale much larger than the thickness
$\Delta R \equiv R_0(u) - R_i(u)$,
its profile is essentially the same as 
that of the ideal thin shell. 

For this configuration \eqref{profile-3}, 
we have
\begin{align}
\psi_0(u) &\simeq
\log\left(\frac{R_i(u)-a_i(u)}{R_i(u)}\right)
- \frac{R_0^2(u) - R_i^2(u)}{4\sigma}
\nn \\
&\simeq
\log\left(\frac{\sigma}{R^2_i(u)}\right)
- \frac{R_0^2(u) - R_i^2(u)}{4\sigma}
\simeq
- \frac{R_0^2(u) - R_i^2(u)}{4\sigma},
\label{psi02}
\end{align}
where in the last line we have assumed that
\begin{align}
\Delta R(u) \equiv
R_0(u) - R_i(u)
\gg \frac{2\sigma}{R_i(u)}\log\left(\frac{R^2_i(u)}{\sigma}\right).
\label{DeltaR}
\end{align}
Notice that,
since $\sigma$ is of the order of the Planck length squared,
the inequality above 
is satisfied even when $\Delta R \equiv R_0(u) - R_i(u)$ is as short as a Planck length
if $R_i^2(u) \gg \sigma$.
This difference $\Delta R(u)$ in the areal radius 
corresponds to a thickness of the shell 
\be
\Delta L \simeq
\sqrt{g_{rr}(R_0(u))} \Delta R(u)
\gg \mathcal{O}\left(\ell_P\log\left(\frac{R_0(u)}{\ell_P}\right)\right).
\label{DeltaL}
\ee
Even for a shell as heavy as $10^{10}$ solar mass,
$\Delta L$ only needs to be greater than $110$ times the Planck length
for the approximation above.
Therefore,
in a low-energy effective theory
applicable only to energies well below $10^{17}$ GeV,
such a pseudo thin shell is
indistinguishable from an ideal thin shell.

To calculate Hawking radiation,
we take the time derivative of eq.\eqref{psi02}:
\begin{align}
\dot{\psi}_0(u)
&\simeq
- \frac{R_0(u)\dot{R}_0(u) - R_i(u)\dot{R}_i(u)}{2\sigma}.
\label{dotpsi2}
\end{align}
Assuming that $R_i(u)$ is falling at the speed of light,
\begin{align}
\dot{R}_i(u)
&= - \frac{1}{2} e^{\psi(u, R_i(u))}\frac{R_i(u)-a_i(u)}{R_i(u)}
\nn \\
&\simeq - \frac{1}{2} e^{- \frac{R_0^2(u) - R_i^2(u)}{4\sigma}}\frac{R_i(u)-a_i(u)}{R_i(u)}.
\end{align}
The huge redshift factor above
\be
e^{- \frac{R_0^2(u) - R_i^2(u)}{4\sigma}}
\simeq e^{-\frac{R_0(u)(R_0(u)-R_i(u))}{2\sigma}}
\gg \frac{\sigma}{R_i^2(u)}
\label{huge-redshift}
\ee
implies that the 2nd term in eq.\eqref{dotpsi2}
is much smaller than the 1st term.
Hence,
\be
\dot{\psi}_0(u)
\simeq
- \frac{R_0(u)\dot{R}_0(u)}{2\sigma}
\simeq
\frac{1}{2a_0(u)},
\label{dotpsi3}
\ee
which is the same as eq.\eqref{dotpsi1}
for a slope-1 shell in Sec.\ref{SmoothConfig},
but differs significantly from the ideal thin shell
in Sec.\ref{idealthinshell-BR}.%
\footnote{%
Eq.\eqref{dotpsi3} differs significantly from 
the ideal thin shell
(with back-reaction from Hawking radiation) in Sec.\ref{idealthinshell-BR}
which has $\dot{\psi}_0\rightarrow 0$.
It also differs by an overall sign 
from the conventional result \eqref{dotpsi0} 
which ignores the back-reaction of Hawking radiation.
}
As a result, 
this thin shell of a finite thickness will evaporate completely,
rather than approaching a classical black hole
like the ideal thin shell.

It is an interesting coincidence that 
the Hawking radiation of the pseudo thin shell
(with its back-reaction included in the calculation)
happen to agree with that of the ideal thin shell
when the back-reaction is ignored,
while they disagree when the back-reaction is turned on.
This observation suggests that,
despite its sensitivity to modifications at a length scale
slightly larger than the Planck scale, 
e.g. $100 \ell_P$ for a shell of $10^{10}$ solar mass,
there is still some robustness in 
the Hawking radiation of the conventional model of black holes.

\subsection{Analogy With Electromagnetism}
\label{LETS}

The pseudo thin shells considered above imply
that the equations governing the Hawking radiation \eqref{Tuu}
is too sensitive to Planck-scale details
so that the notion of ideal thin shells is inappropriate
for discussions in the context of low-energy effective theories.

Similar issues have been discussed in other branches of physics.
It is not uncommon for low-energy effective theories 
to involve higher-derivative terms.
Typically,
such higher-derivative terms lead to various pathologies.
A well known simple example can be found in the 
textbook on classical electromagnetism \cite{Griffiths}.
According to the Abraham-Lorentz formula,
the back-reaction force of the electromagnetic radiation by a point charge
is proportional to the 2nd time-derivative of the velocity of the charge.
Any solution of a point-charge is then either a runaway solution
(accelerating indefinitely to infinite velocity
even after all external forces are removed
--- instability),
or it suffers pre-acceleration
(acceleration before external forces are applied
--- a violation of causality).
The proper way to deal with this issue is,
of course, to keep in mind that,
strictly speaking,
it is unreasonable to claim a charge to be point-like
in a low-energy effective theory.
The equation of motion should be solved for 
a charge with finite size and finite density, 
and the point charge limit, up to the cut-off scale, 
should be taken after solving the equation. 
As long as all charge distributions have a sufficiently smooth profile,
these problems can be ignored.

In the light of this analogy,
a pseudo thin shell satisfying the inequality \eqref{DeltaL})
is a valid thin-shell configuration in the low-energy effective theory,
while the ideal thin shell is not.

It is well known that
the Abraham-Lorentz formula is still applicable
to point charges in perturbative calculations.
At the lowest-order approximation,
a point charge is assumed to move without back-reaction.
The first-order correction to the point-charge trajectory
due to back-reaction can then be calculated,
and it provides a good approximation
whenever the radiation is sufficiently weak.

In general,
in a low-energy effective theory,
one can consider a derivative expansion, 
which would be truncated at a certain order of the expansion. 
Assuming that higher-derivative terms are less important,
the physical result should not be dramatically changed
when the order of the truncation is slightly changed.
However,
higher-derivative terms always introduce new solutions and new instabilities.
An approach to deal with the higher-derivative terms 
in a low-energy effective theory was suggested
by Yang and Feldman \cite{Yang:1950vi},
which was later extended to a general formulation in Ref.\cite{Perturb-Higher}.
Following this prescription,
one can include the effect of higher-derivative terms order by order
without introducing unphysical solutions.
In the rest of this paper, we adopt this approach to take care
of the higher derivatives in the Schwarzian derivative in eq.\eqref{Tuu}.

Notice that,
the ideal thin shell without back-reaction from Hawking radiation
and the pseudo thin shell produces the same formula
\be
\frac{da}{du} \simeq - \frac{\kappa{\cal N}}{48\pi a^2}
\label{dadu-0}
\ee
for Hawking radiation.
This is compatible with our analogy with point charges
in classical electromagnetism.
We will, therefore, take this formula \eqref{dadu-0}
as the lowest-order approximation 
for the Hawking radiation from pseudo thin shells.
Corrections from 
higher-order terms in the Schwarzian derivative
can then be added iteratively order by order. 
We will compute the first-order correction 
and check that it is small in our simulation.

\section{Slope-1 Shell}
\label{SmoothConfig}

In this section,
we review the smooth configuration of the collapsing matter
proposed in Ref.\cite{Kawai:2013mda}:
\be
a(u, r) =
\left\{
\begin{array}{ll}
a_0(u) \equiv R_0(u) - \frac{2\sigma}{R_0(u)}, 
& r > R_0(u),
\\
r - \frac{2\sigma}{r},
& R_0(u) > r > R_1(u),
\\
< r - \frac{2\sigma}{r},
& r < R_1(u),
\end{array}
\right.
\label{smooth-a}
\ee
where $\sigma$ is given by eq.\eqref{sigma}.
We assume that $R_0(u) - R_1(u)$ is of a macroscopic value
much larger than $\frac{\sigma}{a}$.
The matter at $r$ has approached the Schwarzschild radius $a(u,r)$ for 
the total mass inside $r$,
and the slope of $a(u, r)$ as a function of $r$ is approximately equal to $1$
within the range of $r \in (R_1(u), R_0(u))$.
The functional form of $a(u, r)$ within the inner part of the shell for $r < R_1(u)$
will turn out to be irrelevant due to the huge redshift factor.

The slope of the $a-r$ curve is approximately $1$ for this profile,
so we will refer to it as a {\em slope-1 shell}.
The slope-1 shell is interesting because
it is an asymptotic solution compatible with 
the following assumptions:
(1) all layers of the collapsing matter are close to their Schwarzschild radii, and
(2) the decay rate due to Hawking radiation is given by eq.\eqref{da0du-1}.
We also show in the appendix that
slope-1 configurations of an arbitrary thickness
are unique as asymptotic states under certain assumptions.

The outer radius $R_0(u)$ satisfies eq.\eqref{dR0du}
as it falls at the speed of light. 
When the shell is sufficiently close to the Schwarzschild radius,
it can be approximated as 
\begin{equation}
 R_0(u) = a_0(u) - 2 R_0(u) \frac{dR_0(u)}{du} \simeq a_0(u) - 2 a_0(u) \frac{da_0}{du} \ . 
\end{equation}
Combining it with eq.\eqref{da0du-1},
we see that \cite{Kawai:2013mda}
\be
R_0(u) \simeq a_0(u) + \frac{2\sigma}{a_0(u)}.
\label{R0=a0}
\ee
The inverse of this relation is 
\be
a_0(u) \simeq R_0(u) - \frac{2\sigma}{R_0(u)}.
\ee

For spherically symmetric configurations,
we can decompose the collapsing matter into infinitely many 
infinitesimal collapsing layers labeled by a number $n$.
The total mass enclosed in each shell of radius $R_n$ defines
the Schwarzschild radius $a_n$ associated to that shell.
(The geometry of the infinitesimal gap between 
the $n$-th layer and the $(n+1)$-st layer is thus determined by $a_n$.)
we can, therefore, apply the same argument 
to every layer below the surface,
to claim that eventually 
\be
R_n(u) \simeq a_n(u) + \frac{2\sigma}{a_n(u)},
\ee
and hence eq.\eqref{smooth-a} is motivated.
This was why the slope-1 configuration was referred to
as the ``asymptotic black hole'' in Ref.\cite{Ho:2016acf}.

For this smooth configuration \eqref{smooth-a},
eq.\eqref{psi0} implies that
\be
\psi_0(u) \simeq - \frac{R_0^2(u)}{4\sigma},
\ee
so that
\be
\dot{\psi}_0(u) \simeq - \frac{R_0(u)\dot{R}_0(u)}{2\sigma}
\simeq - \frac{a_0(u)\dot{a}_0(u)}{2\sigma}
\simeq \frac{1}{2a_0(u)}.
\label{dotpsi1}
\ee
Notice that this differs from the conventional result \eqref{dotpsi0}
for the ideal thin shell without back-reaction by a sign,
but the Hawking radiation is the same at the leading order!
It is very different from the case of the ideal thin shell 
when the back-reaction of Hawking radiation is taken into consideration.

It was proposed in Ref.\cite{Ho:2016acf} that,
from the viewpoints of distant observers, 
the slope-1 configuration is expected to appear as 
an asymptotic configuration 
for generic initial states of the collapsing matter.
On the other hand, 
it was also noted there that,
as the surface of the collapsing shell
gets very close to the Schwarzschild radius
(i.e. when eq.\eqref{R0=a0} is satisfied),
everything below a short depth under the surface
is essentially frozen.
It is, therefore, possible that 
the collapsing matter demonstrates a different asymptotic profile.
One of the main results of this paper is to show that, 
even though the slope-1 configuration is indeed an asymptotic state,
there exists another asymptotic state.

\section{Asymptotic States}
\label{AS}

The rest of the paper is focused on the question
``What are the asymptotic states of gravitational collapses?''
First,
in Sec.\ref{StaticAsympt},
we consider the collapsing layers in the static Schwarzschild background. 
The back-reaction of Hawking radiation is ignored,
and the geometry between the collapsing layers is given by 
the static Schwarzschild background.
In Sec.\ref{TDAsympt},
we describe our result about
how things are different when the back-reaction is turned on.

\subsection{Without Back-Reaction of Hawking Radiation}
\label{StaticAsympt}

For a static Schwarzschild background,
each layer in the collapsing matter approaches its event horizon.
If the initial profile has a slope $da/dr$ much smaller than $1$,
the inner layers approach their Schwarzschild radii earlier,
and the outer layers to theirs later.
The inner layers do not impose redshift factors
to slow down the collapse of outer layers for a distant observer.
Therefore,
in the end,
for distant observers,
all the layers are quite close to their Schwarzschild radii
and the profile of the whole collapsing matter is frozen.
Such configurations approach
states with slopes $da/dr$ approximately equal to $1$,
although in detail they are in general different from
the slope-1 state described in Sec.\ref{SmoothConfig}.
The attractor states have slope $1$ for initial states
with sufficiently small slopes $da/dr$.

If, however,
the initial state has a slope $da/dr$ larger than $1$,
the outer layers approach their Schwarzschild radii earlier,
and they impose a large redshift factor on the inner layers
from the viewpoint of a distant observer.
With the inner layers frozen by the large redshift factor,
the profile of the inner layers remains essentially the same as the initial state.
There is no unique asymptotic profile in this case.

This picture will be modified
by turning on the back-reaction of Hawking radiation.

\subsection{Back-Reaction of Hawking Radiation}
\label{TDAsympt}

When the back-reaction of Hawking radiation 
is taken into consideration,
the situation is slightly changed. 
The geometries between the layers 
are given by the outgoing Vaidya metric, 
whose Schwarzschild radii have time dependence and are decreasing 
due to the effect of Hawking radiation. 
The profile of the energy distribution is affected by 
the time dependence of the Schwarzschild radii between the layers. 

For a profile with $da/dr \ll 1$,
the inner layers would get close to their Schwarzschild radii
before the outer layers do theirs,
so the inner layers evaporate first.
As in the case without the back-reaction,
the outer layers are never frozen,
so they keep falling in
until they also approach their Schwarzschild radii.
Eventually, all layers are close to their Schwarzschild radii,
and this is the slope-1 configuration (Sec.\ref{SmoothConfig}).
It looks approximately the same as the asymptotic state
without the back-reaction in Sec.\ref{StaticAsympt}.
The effect of the back-reaction gives no significant modification 
before the layers approach the slope-1 configuration. 
The layers are still moving inward even after 
it becomes the slope-1 configuration since 
the Schwarzschild radii are decreasing. 
However,
the profile remains the same.

On the other hand,
for an initial state with a large slope $da/dr \gg 1$,
the outer layers reach close to their Schwarzschild radii $a(u, r)$
before the inner layers do.
The inner layers are frozen by the $1/(r-a)$ contribution
of the outer layers due to the redshift factor \eqref{psiur},
and the outer layers evaporate first.
As the outer layers are evaporating,
the thickness of an outer layer close to its Schwarzschild radius $a(u, r)$
reduces over time until it is barely thick enough to keep the inner shells frozen.
As outer layers evaporate away,
inner layers are thawed and start falling close to their Schwarzschild radii.
After the evaporation of the outer layers, 
the outermost part of the inner layers plays the role of the outer layers. 
If the profile of the inner layers also has a large slope $da/dr \gg 1$, 
the same process to the above is repeated. 
Thus, the thickness of the whole shell decreases over time.
Although the inner part of the collapsing matter deep under its surface
still has an arbitrary profile depending on the initial condition,
the part close to the surface approaches a thin shell.
The thin shells can thus be viewed as another class of asymptotic states.
The back-reaction of Hawking radiation plays an important role in 
this mechanism.
This result is quite different from the case without the back-reaction.

\section{Numerical Simulation}
\label{Numerical}

In this section,
through numerical methods,
we investigate the dynamical process
leading to the asymptotic states
described in the previous section.
As it was explained in Sec.\ref{ThinShellSec},
we shall not consider ideal thin shells
and the higher-derivative terms in the Hawking radiation
should be treated in a way suitable for
the low-energy effective theory.

\subsection{Numerical Methods}

In numerical simulation,
we discretized the continuous matter distribution 
as a set of collapsing thin shells
labeled by a 
number $n = 1, 2, \cdots, N$ from the innermost shell ($n=1$)
to the outermost shell ($n=N$).
Each shell has a radius $R_n$,
and the total mass $M_n$ enclosed in the shell defines 
the Schwarzschild radius $a_n = 2G M_n$ associated with the shell.
The metric between the $n$-th shell and the $(n+1)$-st shell
is given by the outgoing Vaidya metric and determined by the Schwarzschild radius $a_n$.

As it was explained in Sec.\ref{LETS},
we adopt the perturbative approach
in which each layer of the thin shell evaporates according to eq.\eqref{dadu-0},
to avoid unphysical dependence on Planck-scale structures.
Hence,
we assume that for the $n$-th layer,
at the leading order
\be
\frac{da_n}{du_n} \simeq - \frac{\sigma}{a_n^2}
\qquad
\left(\sigma \equiv \frac{\kappa{\cal N}}{48\pi}\right),
\label{dandun}
\ee
where $u_n$ is the outgoing light-cone coordinate
for the segment between the $n$-th and the $(n+1)$-st shell.
This assumption is 
iterated back into
the Schwarzian derivative $\{u_n, U\}$
for the first-order correction.

Schematically,
the first-order correction is calculated in the following way.
The Schwarzian derivative introduces first and second-order derivatives 
of $a_n$ into the decay rate equation of $a_n$ as
\be\label{eq:30}
\begin{split}
\frac{da_n}{du_n}
=
f\left(a_n, \frac{da_n}{du_n}, \frac{d^2 a_n}{du_n^2}, R_n\right),
\end{split}
\ee
through a certain function $f$.
In general,
there could be dependence on the first and second derivatives of $R_n$
in the function $f$,
but they can be traded for zeroth and first derivatives of $a_n$
through the evolution equation of $R_n$
\be
\frac{dR_n}{du_n} = - \frac{1}{2}\left(1 - \frac{a_n}{R_n}\right) , 
\ee
respectively. 
By using this equation repeatedly, the derivatives of $R_n$ can be removed from 
the expression of $f$, and therefore, $f$ in \eqref{eq:30} does 
not depend on the derivatives of $R_n$. 
Thus the decay rate equation \eqref{eq:30} contains 
only the derivatives of $a_n$. 

As we explained in Sec.\eqref{LETS},
to avoid the pathologies introduced by higher derivatives, 
we will solve the equation by the perturbative method.
We first take the conventional formula \eqref{dandun}
as the zeroth-order approximation for $\frac{da_n}{du_n}$.
As a result,
\be\label{eq:9}
 \frac{d^{2}a_n}{du_n^{2}}
 \simeq \frac{-2\sigma^2}{a_n^5},
\ee
so that $\frac{da_n}{du_n}$ and $\frac{d^2 a_n}{du_n^2}$
on the right-hand side of eq.\eqref{eq:30}
can be replaced by functions of $a_n$
and $R_n$, 
without their time derivatives at all.
This way, we get a first-order differential equation for $a_n$. 
That is, $\frac{da_n}{du_n}$ is given by a function of $a_n$ and $R_n$.
(It is straightforward to derive this lengthy expression
so we will not present it here.)
In general, 
one can do this iteratively to get higher-order corrections
to obtain a more accurate approximation.
We will check that the first-order correction is small,
and ignore higher-order corrections.

To study the astrophysical black holes,
we need to consider a sufficiently large mass. 
To study such a large mass, a huge number of layers are necessary. 
However, the simulation of a lot of layers is computationally expensive. 
Here, we are interested in the asymptotic states of the collapsing layers, 
in particular, whether they approach the slope-1 state
or evaporate from the outer layers. 
To see this, it is sufficient to study the outer shells. 
To save computer time in our simulation, 
we sometimes include a massive core at the center
that also evaporates according to the conventional formula \eqref{dandun}. 
This way we can efficiently describe a configuration with a large mass,
while we focus on the behaviors of the outer shells,
and see if they evaporate first.
Note that the inner core is essentially frozen 
because of the strong redshift factor due to the outer shells
when the outer shells are close to their Schwarzschild radii,
so we do not expect the replacement of many inner shells by a massive core
to make much difference to the behavior of the outer shells.
To verify this assumption,
one can compare the dynamics of the outer $n$ layers
of a system of $N+n$ layers of collapsing shells,
versus the dynamics of $n$ layers in a second system,
in which a massive core replaces the $N$ shells of the first system.
One can check that the dynamics of the $n$ shells in both systems agree.
Furthermore,
one can check that the geometry outside all the shells 
agrees with that of a single massive core of the same total mass.
Hence the assumption can be justified by induction.

\subsection{Numerical Results}
\label{NumRes}

Now, we study the time evolution of collapsing layers, 
taking the back-reaction of Hawking radiation into account. 
Unless otherwise specified, 
the time evolution $R_n$ and $a_n$ will be presented
as functions of the light-like coordinate $u=u_0$ outside all shells.

We assume that there is no horizon in the initial state,
that is, no shell is inside its Schwarzschild radius, 
\be
R_{n}(u_\text{init}) > a_{n}(u_\text{init})
\qquad \forall n
\ee
at the initial time $u = u_\text{init}$. 
$R_{n}(u)$ decreases as the $n$-th shell collapses. 
Its associated Schwarzschild radius $a_{n}(u)$
slowly decreases due to Hawking radiation
when $R_{n}(u)$ gets close to $a_{n}(u)$.

The profile of a collapsing matter distribution at any given time $u$ 
is given by the plot of $a_{n}$ vs $R_{n}$ for that $u$.
While the profile (the $a-r$ graph) of the initial configuration 
is arbitrary except that it must be a monotonically increasing relation,
we shall focus on linear profiles with various slopes.
Once they are understood,
it is straightforward to
generalize the knowledge to
a generic profile 
by decomposing the profile into
many small segments which are approximated
by linear relations between $a$ and $r$.

With the assumption that
all layers $n$ are falling at the speed of light,
the radius $R_{n}$ satisfies the equation
\be
\frac{dR_{n}(u)}{du} = 
- \frac{1}{2} e^{\psi_{n}}\frac{R_{n} - a_{n}}{R_{n}}.
\ee
This assumption also ensures that the shells do not cross each other. 
When the collapsing matter is still far away from its Schwarzschild radius,
at large distances where the spacetime is nearly flat,
the $a-r$ profile remains roughly unchanged 
as it shifts along the $r$-axis over time.
We shall, therefore, focus on the initial states
in which at least a part of the $a-r$ profile 
is very close to the $a = r$ line.
That is, some of the layers have already approached 
their Schwarzschild radius | 
$R_{n} \simeq a_{n}$ for at least some values of $n$.

\begin{figure}[h]
\includegraphics[keepaspectratio, width=9cm, height=7.5cm]{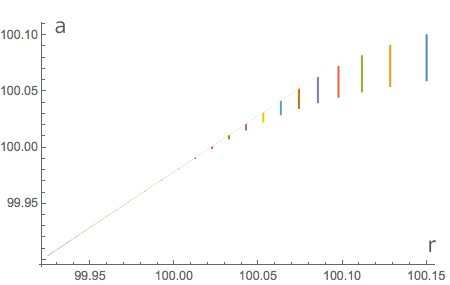} 
\caption{
$a-r$ graph for $\frac{\del a}{\del r} = \infty$ \\
The profile of the collapsing sphere is shown at different instants of $u$.
The initial state with $\frac{\del a}{\del r} = \infty$
is the first vertical line on the right.
The profile gets shorter as it moves to the left
due to evaporation.
The slope remains infinite throughout the dynamical process.
}
\label{fig:rad0}
\end{figure}

\begin{figure}[h]
\includegraphics[keepaspectratio, width=9cm, height=7.5cm]{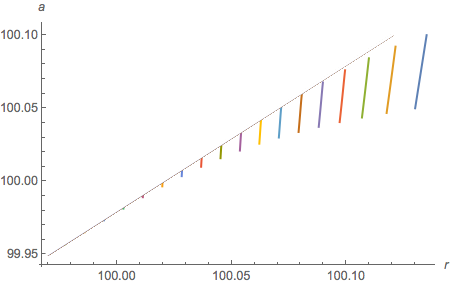} 
\caption{
$a-r$ graph for $\frac{\del a}{\del r} = 10$ \\
The profile of the collapsing sphere is shown at different instants of $u$.
The initial state with $\frac{\del a}{\del r} = 10$
is the first vertical line on the right.
The profile gets shorter as it moves to the left
due to evaporation.
The slope approaches infinite in the dynamical process.
}
\label{fig:rad00}
\end{figure}

\begin{figure}[h]
\includegraphics[keepaspectratio, width=9cm, height=7.5cm]{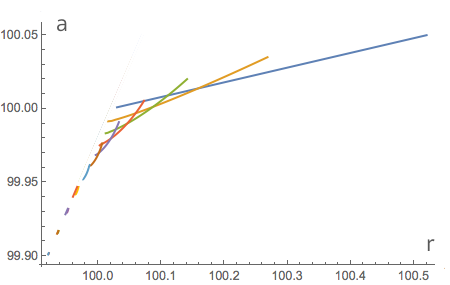} 
\caption{
$a-r$ graph for $\frac{\del a}{\del r} = 0.1$ \\
The profile of the collapsing sphere is shown at different instants of $u$.
The initial state with $\frac{\del a}{\del r} = 0.1$
is the first straight line (blue).
The profile gets shorter and the slope becomes larger
as it moves to the left.
Eventually, the slope approaches $1$.
}
\label{fig:rad01}
\end{figure}

\begin{figure}[h]
\includegraphics[keepaspectratio, width=9cm, height=7.5cm]{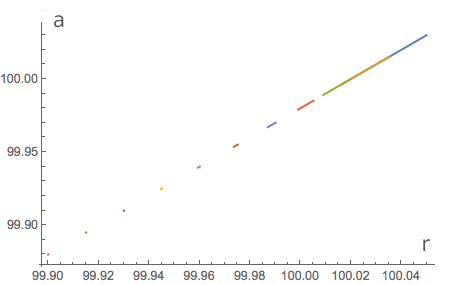} 
\caption{
$a-r$ graph for $\frac{\del a}{\del r} = 1$ \\
The profile of the collapsing sphere is shown at different instants of $u$.
The initial state with $\frac{\del a}{\del r} = 1$
is the diagonal line in blue,
partially overlapping with the profile at later times.
The profile gets shorter and the slope remains approximately equal to $1$.
}
\label{fig:rad1}
\end{figure}

When the radius $R_{n}$ of a layer $n$ is close 
to its Schwarzschild radius $a_{n}$,
the decreasing rate in $R_{n}$ slows down. 
If the initial slope is smaller than 1, 
the innermost layers get close to the curve $a=r$ first, 
and then, the slope $\del a/\del r$ increases over time.
If the initial slope is larger than 1, 
the outermost layers approach $a=r$ first. 
In this case, the inner layers also slow down because of 
the redshift factor due to the outer shells. 
Therefore, the slope does not decrease. 
See Figs.\ref{fig:rad0}--\ref{fig:rad1} for the profiles at each moment of the collapse.
It should be noted that, for a large initial slope of the profile, 
the effect of the redshift factor appears 
because the time evolution is measured by using the time coordinate outside all shells. 
Although the collapse does not slow down for the local observer 
at the inner layers,
it slows down for
the observer outside the collapsing matters. 

Despite the tendency of increasing the slope $\del a/\del r$ over time,
once a layer is very close to the $a = r$ line,
it can only move diagonally along the $a = r$ line
(more precisely,
a curve along which $r \simeq a + \frac{2\sigma}{a}$).
Our simulation shows that,
for initial configurations with slopes $\del a/\del r \ll 1$,
the slope only approaches $1$ in the end,
while configurations with initial slopes $\del a/\del r \gg 1$, 
the slope approaches infinity,
which implies that the collapsing layers approach a thin shell. 


To conclude,
there are two distinct classes of 
asymptotic states of black holes. 
The first class approaches the slope-$1$ configuration,
for initial profiles with a small slope.
The second class approaches thin-shell configurations
for initial profiles with large slopes.
The criterion deciding whether an initial configuration
evolves towards one class or another is 
whether the slope of the initial profile $da/dr$ is much smaller than $1$
or much larger than $1$.

We considered only the profiles with constant slopes so far. 
It is straightforward to study more generic profiles. 
For example, if the slope for inner layers is large 
but that for outer layers is small, 
the middle layers
will approach the line of $a=r$ first. 
The inner layers will be frozen and moving inward without changing the slope, 
while the outer layers will asymptote to the slope-1 configuration. 

Recall that the volume density of matter is defined as
\be
\frac{dm}{d\mbox{vol}} = \frac{1}{8\pi G_N r^2}\frac{da}{dr},
\ee
so the critical volume density $\rho_c$
(around the Schwarzschild radius)
corresponding to $da/dr = 1$ is
\be
\rho_c \equiv 
\left.\frac{1}{4\pi r^2}\frac{\del M}{\del r}\right|_{\mbox{\tiny critical}} 
\simeq \frac{1}{8\pi G_N a^2} = \frac{1}{\kappa a^2},
\label{rho-c}
\ee
where we have used
$a = 2G_N M$.
This corresponds to a volume density of order $\mathcal{O}(1/a^4)$
at a distance of order $\mathcal{O}(a)$ away from the horizon.
It is of the same order of magnitude as the Hawking radiation.
In fact,
the ingoing energy of the collapsing matters 
balances with that of Hawking radiation 
in the special case of the slope-1 configuration
which is studied in \cite{Kawai:2013mda}. 
Hence,
we expect the ingoing energy flux of density higher (lower) than Hawking radiation
to resemble the thin shell (slope-1 configuration) 
as it approaches the horizon.

Even for a black hole as small as a solar mass,
its Hawking radiation is much weaker than 
the current cosmic microwave background radiation.
Hence all existing black holes are expected to have a surface layer
resembling the thin shell configuration.
That is,
the slope $da/dr$ is very large at the surface of the black hole.
%

\subsection{Stages of Evaporation}

In the KMY model,
it was found \cite{Kawai:2013mda} that,
for the ideal slop-$1$ shells,
the collapsing shells are moving inward as 
they lose the energy by Hawking radiation.
It is also argued \cite{Kawai:2015uya} that 
the evaporation of the collapsing shells, in most cases, 
happens in the same fashion as peeling an onion ---
layer by layer from the outside.
The reason is that
the inner shells are frozen due to the huge redshift factor.

Here we would like to ask 
in general, when would
the evaporation start from the layers on the outside or the inside.
To see this, we calculate the time evolution of 
the Schwarzschild radii for each shell, as functions of $u$. 
We find that the evaporation process is a competition
between the suppression due to a large redshift factor
and the enhancement of evaporation due to
the short separation of a layer from its Schwarzschild radius $a$.
The details are described in the following.

\subsubsection{Small-Slope Configurations}

For the profile of a collapsing matter with a very small slope ($da/dr \ll 1$),
the innermost shell approaches its Schwarzschild radius
before the other shells and starts to evaporate first.
As more and more layers approach their Schwarzschild radii,
a layer that is already close to its Schwarzschild radii
will stay locked at a separation $R_n - a_n \simeq 2\sigma/a_n$ 
until it is evaporated.

As the outer layers approach their Schwarzschild radii,
the redshift factor becomes large for the remaining inner layers,
so that the radiation from the remaining inner layers is suppressed.
The outer layers are then evaporated before them.

Therefore,
for a profile with a tiny slope,
its evaporation can be very roughly decomposed into two stages.
In the first stage,
the inner layers evaporate first,
and the outer layers are still far from their Schwarzschild radii.
In the 2nd stage%
\footnote{%
This is the stage of onion peeling 
described in the KMY model. 
},
after it has turned into a slope-1 configuration,
the outer layers get close to their Schwarzschild radii
and start to evaporate,
and the inner layers surviving the first stage are frozen
until outer layers are evaporated.
The configuration of the collapsing matter is 
now in agreement with the slope-1 shell.

\begin{figure}[h]
\includegraphics[keepaspectratio, width=7.5cm, height=7.5cm]{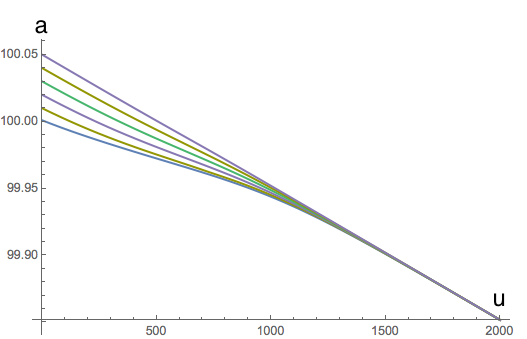} 
\caption{
$a(u)$ for $da/dr = 0.1$ \\
The values $a_n$ for all layers are shown as functions of time $u$.
The values of $a_n$ do not start at $0$
because there is a massive core at the center.
The innermost layers evaporate first,
so in the diagram,
they merge with other layers
initially on top of them.
At larger $u$,
the outermost layers evaporate
and appear to merge with other layers initially under them.
}
\label{fig:aud01}
\end{figure}

\begin{figure}[h]
\includegraphics[keepaspectratio, width=7.5cm, height=7.5cm]{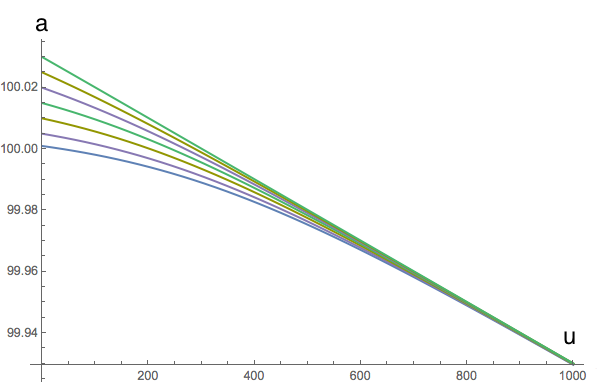}
\caption{
$a(u)$ for $da/dr = 1$ \\
The values $a_n$ for all layers are shown as functions of time $u$.
The values of $a_n$ do not start at $0$
because there is a massive core at the center.
The outermost layers evaporate first
so they appear to merge with other layers initially under them.
}
\label{fig:aud1}
\end{figure}

This process is shown in Fig.\ref{fig:aud01}.
The plot consists of many curves of the Schwarzschild radii for each shell. 
A layer completely evaporates when the curve merges with the next curve | 
its Schwarzschild radius becomes the same as that of the inner layer. 
In comparison,
the process for $\del a/\del r \simeq 1$ is shown in Fig.\ref{fig:aud1},
for which the outer layers are evaporated first,
resembling the late stage of Fig.\ref{fig:aud01}.

\begin{figure}[h]
\includegraphics[keepaspectratio, width=11.25cm, height=7.5cm]{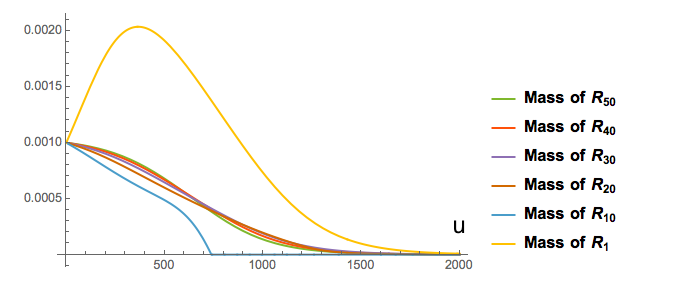} 
\caption{
$m(u)$ for $\del a/\del r = 0.1$ \\
The masses of some of the layers are shown as functions of time $u$.
It is clear that, generically,
inner layers evaporate faster in the beginning,
while outer layers evaporate faster at a later time.
There is roughly a point of intersection of the curves
when they have evaporated the same percentage of their masses.
The mass of the first layer $R_1$ outside the massive core
appears to be anomalous as it initially increases with time.
This is because the outgoing energy due to the evaporation
of the massive core is temporarily counted as its energy.
This is merely an artifact of our choice of the configuration.
}
\label{fig:evad01}
\end{figure}

\begin{figure}[h]
\includegraphics[keepaspectratio, width=11.25cm, height=7.5cm]{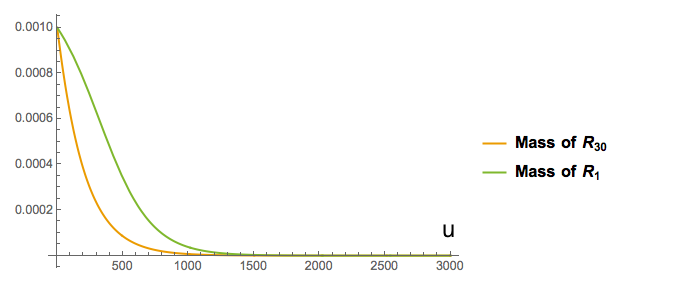}
\caption{
$m(u)$ for $\del a/\del r = 1$ \\
The masses of some of the layers are shown as functions of time $u$.
The situation is simple in this case:
outer layers evaporate faster than inner layers at all times.
}
\label{fig:evad1}
\end{figure}


\begin{figure}[h]
\begin{subfigure}{0.4\textwidth}
\includegraphics[scale=0.4]{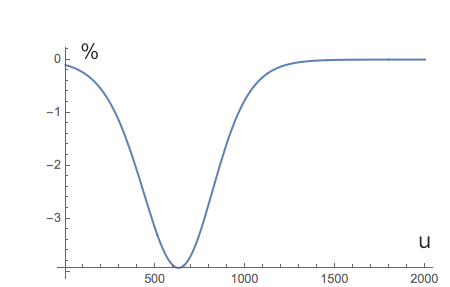} 
\caption{
The first-order correction $\Delta\left(\frac{da}{du}\right)$
as percentage of the 0th-order expression
of $\frac{da}{du}$.
}
\label{fig:dareladiffd001}
\end{subfigure}
\begin{subfigure}{0.4\textwidth}
\includegraphics[scale=0.4]{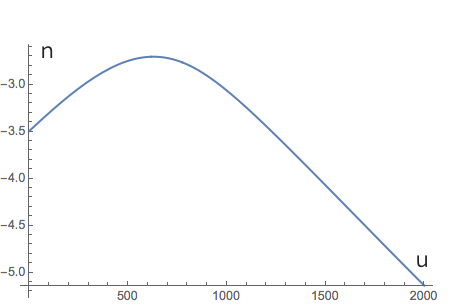} 
\caption{
Modeling the dependence of the first-order correction 
$\Delta\left(\frac{da}{du}\right)$ by power law 
$\Delta\left(\frac{da}{du}\right) = A a^n$
for some constant $A$.
}
\label{fig:daorderd001}
\end{subfigure}
\begin{subfigure}{0.4\textwidth}
\hskip2em
\includegraphics[scale=0.4]{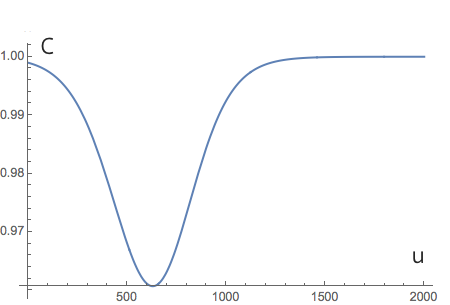} 
\caption{
Modeling the first-order correction by the coefficient $C$ in
$\Delta\left(\frac{da}{du}\right) = \frac{(1-C)\sigma}{a^2}$.
}
\label{fig:daconstantd001}
\end{subfigure}
\caption{
1st-order correction in Hawking radiation for $\frac{da}{dr} = 0.1$
}
\label{fig:da001}
\end{figure}

\begin{figure}[h]
\begin{subfigure}{0.4\textwidth}
\includegraphics[scale=0.4]{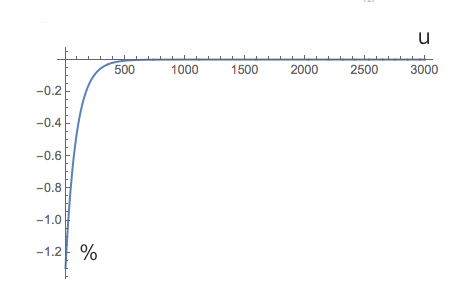}
\caption{
The first-order correction $\Delta\left(\frac{da}{du}\right)$
as percentage of the 0th-order expression
of $\frac{da}{du}$.
}
\label{fig:dareladiffd1}
\end{subfigure}
\begin{subfigure}{0.4\textwidth}
\includegraphics[scale=0.4]{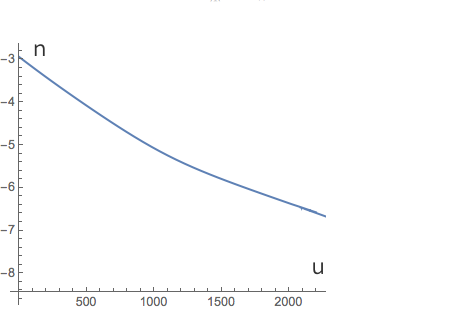}
\caption{
Modeling the dependence of the first-order correction 
$\Delta\left(\frac{da}{du}\right)$ by power law 
$\Delta\left(\frac{da}{du}\right) = A a^n$
for some constant $A$.
}
\label{fig:daorderd1}
\end{subfigure}
\begin{subfigure}{0.4\textwidth}
\hskip2em
\includegraphics[scale=0.19]{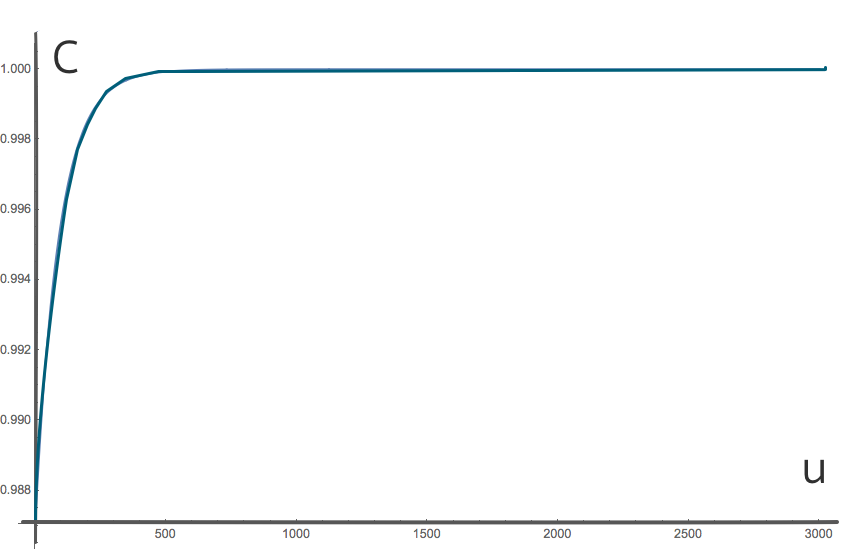}
\caption{
Modeling the first-order correction by the coefficient $C$ in
$\Delta\left(\frac{da}{du}\right) = \frac{(1-C)\sigma}{a^2}$.
}
\label{fig:daconstantd1}
\end{subfigure}
\caption{
1st-order correction in Hawking radiation for $\frac{da}{dr} = 1$.}
\label{fig:da1}
\end{figure}

By the analogy with the radiation of point charges in classical electromagnetism
in Sec.\ref{LETS}, 
the gravitational collapse with Hawking radiation in our formulation
should be free of pathological instabilities.
Indeed, 
in all of our numerical simulation,
the system under study always approaches one of the asymptotic states.
Nevertheless,
we study in more detail the transition process towards the asymptotic states.

The time-dependence of the masses of different shells are plotted
in Figs.\ref{fig:evad01} and \ref{fig:evad1} for $\del a/\del r = 0.1$ and $\del a/\del r = 1$,
respectively.
In both cases,
after a short time with non-uniform behavior,
the collapsing shells approach the asymptotic states.

For both $\frac{da}{dr}$ equal to $0.1$ and $1$,
the first-order correction to the 0th-order Hawking radiation
is relatively small. 
We analyzed this deviation in three different ways:
(1) the percentage of correction with respect to the 0th-order amplitude:
$\Delta\left(\frac{da}{du}\right)/\frac{da^{(0)}}{du}$,
(2) how the correction $\Delta\left(\frac{da}{du}\right)$ scales with $a$
during the collapsing process,
and 
(3) If we use the formula
$\frac{da}{du} = - C(u)\sigma/a^2$ to model the Hawking radiation,
how is the coefficient $C(u)$ changing with time.
In all analyses,
we find the first-order correction 
to Hawking radiation sufficiently small to justify
our perturbative interpretation of the Schwarzian derivative. 
See Fig.\ref{fig:da001} for $\frac{da}{dr} = 0.1$
and Fig.\ref{fig:da1} for $\frac{da}{dr} = 1$.

\subsubsection{Large-Slope Configurations}

For a collapsing profile with a large slope ($da/dr \gg 1$),
the outer layers reach close to their Schwarzschild radii earlier
than the inner layers.
Hence the inner layers are frozen until outer layers are evaporated.

\begin{figure}[h]
\includegraphics[keepaspectratio, width=7.5cm, height=7.5cm]{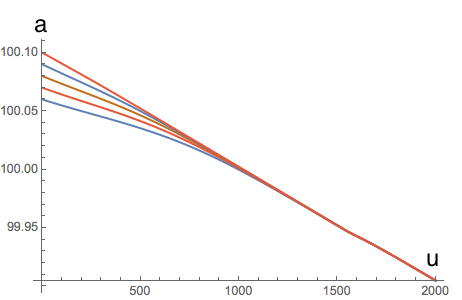}
\caption{
$a(u)$ for $\del a/\del r = \infty$ \\
The values $a_n$ for all layers are shown as functions of time $u$.
The values of $a_n$ do not start at $0$
because there is a massive core at the center.
The outermost layers evaporate first,
so in the diagram, they appear to merge with other layers
initially below them.
}
\label{fig:aud0}
\end{figure}

\begin{figure}[h]
\includegraphics[keepaspectratio, width=7.5cm, height=7.5cm]{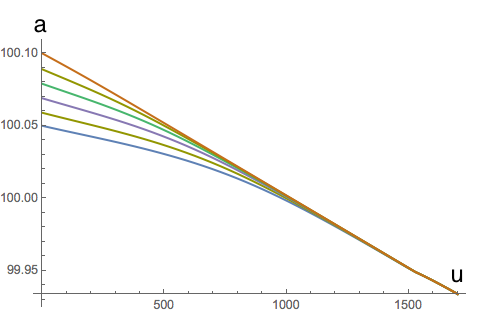}
\caption{
$a(u)$ for $\del a/\del r = 10$ \\
This diagram is essentially the same as Fig.\ref{fig:aud0}.
}
\label{fig:xxx1}
\end{figure}

Whenever the outer layers are very close to their Schwarzschild radii,
the inner layers are frozen by the large redshift factor,
and the radiation of the inner layers can be ignored
(in terms of the time coordinate $u$)
regardless of whether the inner layers are close to 
their Schwarzschild radii.
Therefore, 
for a collapsing ball with an outer surface
that has been evaporating for a long time,
it is evaporating from the outer layers in most cases.
See Figs.\ref{fig:aud0} and \ref{fig:xxx1}.

An exception would be 
the stationary solution which is studied in \cite{Kawai:2013mda}. 
In this case,
the inner layers are also very close to their Schwarzschild radii 
and the evaporation proceeds simultaneously. 
On the other hand,
even in this case,
although the inner layers completely evaporate first, 
the process is extremely slow because of the very large redshift factor, 
and the Hawking radiation mostly comes from the outer layers.

The inner layers stay frozen until the outer layers are evaporated.
For an outer layer as close to the Schwarzschild radius
as $R_n - a_n \sim \mathcal{O}(\sigma/a_n)$,
the redshift factor for the layer at a separation $\Delta R$
under the outer surface is of order $e^{-R_0\Delta R/2\sigma}$.
Therefore,
roughly speaking,
it is natural to think of a layer of matter
to have the thickness of a Planck length,
and the layer is evaporated before the next layer starts to evaporate.

\begin{figure}[h]
\includegraphics[keepaspectratio, width=11.25cm, height=7.5cm]{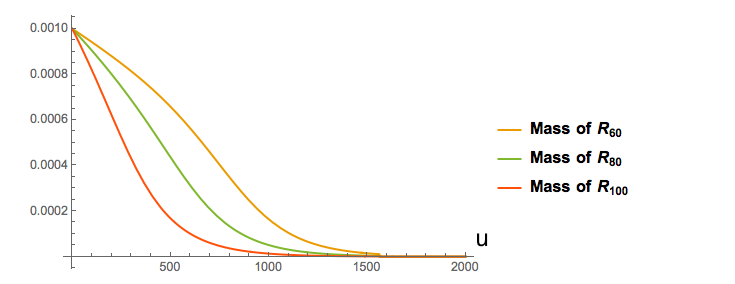}
\caption{
$m(u)$ for $\del a/\del r = \infty$ \\
The masses of some of the layers are shown as functions of time $u$.
Generically,
outer layers evaporate faster.
}
\label{fig:evad0}
\end{figure}

\begin{figure}[h]
\hskip0.4cm
\includegraphics[keepaspectratio, width=9.95cm, height=7.5cm]{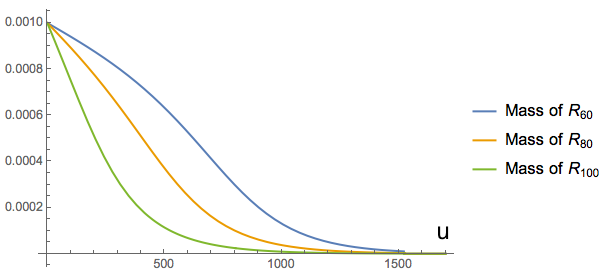}
\caption{
$m(u)$ for $\del a/\del r = 10$ \\
This diagram is essentially the same as Fig.\ref{fig:evad0}.
}
\label{fig:xxx2}
\end{figure}


\begin{figure}[h]
\begin{subfigure}{0.4\textwidth}
\includegraphics[width=0.9\linewidth, height=5cm]{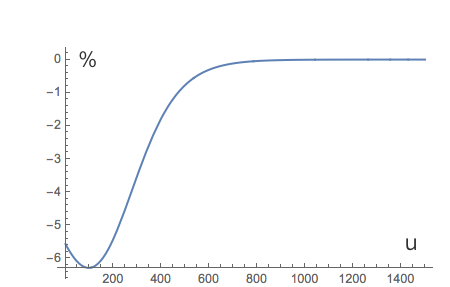}
\caption{
The first-order correction $\Delta\left(\frac{da}{du}\right)$
as percentage of the 0th-order expression
of $\frac{da}{du}$.
}
\label{fig:dareladiffd0}
\end{subfigure}
\begin{subfigure}{0.4\textwidth}
\includegraphics[scale=0.4]{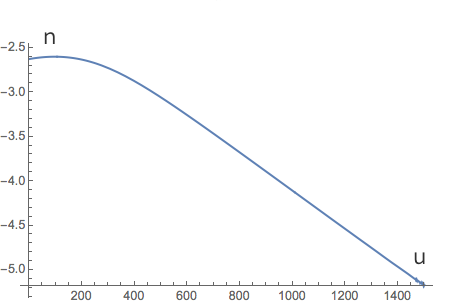}
\caption{
Modeling the dependence of the first-order correction 
$\Delta\left(\frac{da}{du}\right)$ by power law 
$\Delta\left(\frac{da}{du}\right) = A a^n$
for some constant $A$.
}
\label{fig:daorderd0}
\end{subfigure}
\begin{subfigure}{0.4\textwidth}
\includegraphics[scale=0.4]{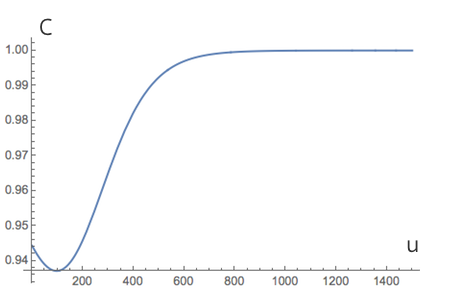}
\caption{
Modeling the first-order correction by the coefficient $C$ in
$\Delta\left(\frac{da}{du}\right) = \frac{(1-C)\sigma}{a^2}$.
}
\label{fig:daconstantd0}
\end{subfigure}
\caption{
1st-order correction in Hawking radiation for $\frac{da}{dr} = \infty$
}
\label{fig:da0}
\end{figure}

\begin{figure}[h]
\begin{subfigure}{0.4\textwidth}
\includegraphics[width=0.9\linewidth, height=5cm]{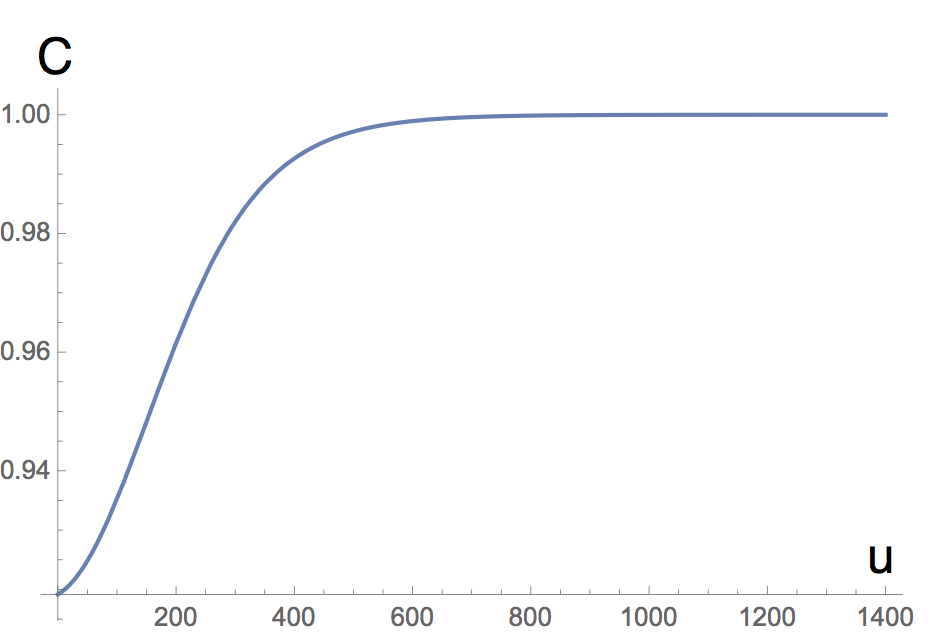}
\caption{
The first-order correction $\Delta\left(\frac{da}{du}\right)$
as percentage of the 0th-order expression
of $\frac{da}{du}$.
}
\label{fig:dareladiffd0}
\end{subfigure}
\begin{subfigure}{0.4\textwidth}
\includegraphics[scale=0.4]{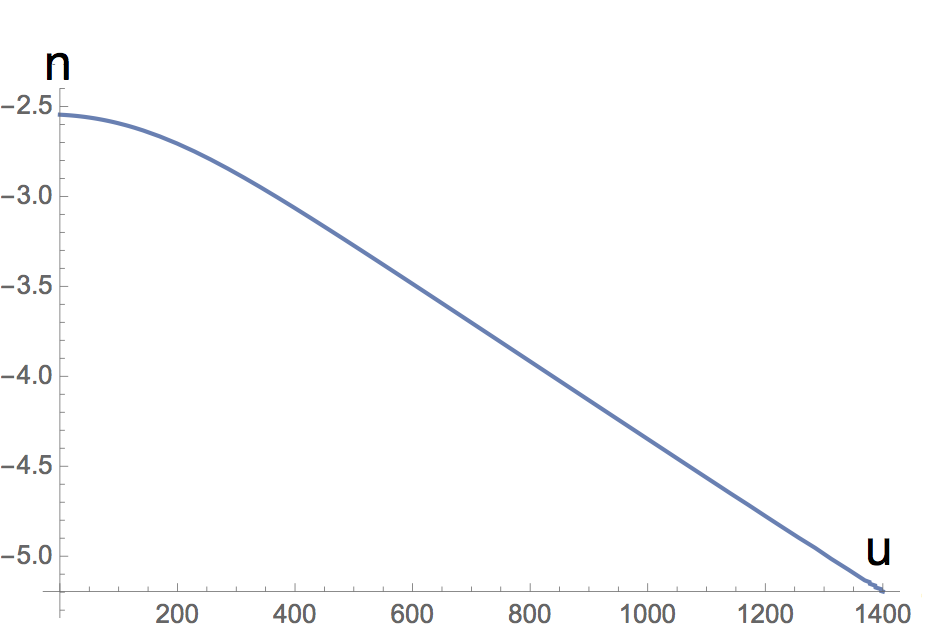}
\caption{
Modeling the dependence of the first-order correction 
$\Delta\left(\frac{da}{du}\right)$ by power law 
$\Delta\left(\frac{da}{du}\right) = A a^n$
for some constant $A$.
}
\label{fig:daorderd0}
\end{subfigure}
\begin{subfigure}{0.4\textwidth}
\includegraphics[scale=0.4]{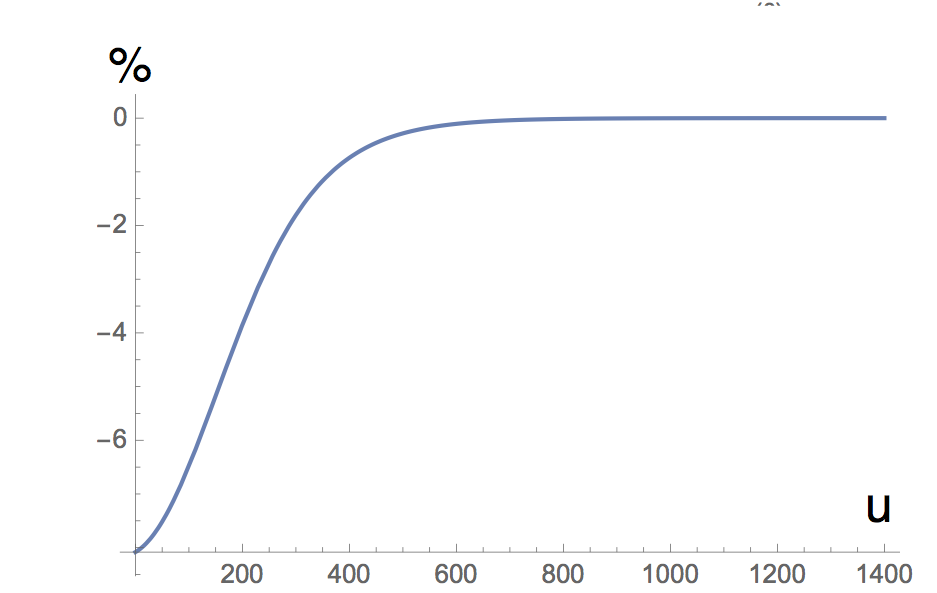}
\caption{
Modeling the first-order correction by the coefficient $C$ in
$\Delta\left(\frac{da}{du}\right) = \frac{(1-C)\sigma}{a^2}$.
}
\label{fig:daconstantd0}
\end{subfigure}
\caption{
1st-order correction in Hawking radiation for $\frac{da}{dr} = 10$
}
\label{fig:dadr=10}
\end{figure}

We also check the issue of stability for the large-slope configurations
as we did for the small-slope configurations above.
The transition process towards the asymptotic states is
shown via the time-dependence of the masses of different shells.
See Figs.\ref{fig:evad0} and \ref{fig:xxx2}.
For both $\frac{da}{dr} = \infty$ and $\frac{da}{dr} = 10$,
the first-order correction to the 0th-order Hawking radiation
is also very small
(see Figs.\ref{fig:da0} and \ref{fig:dadr=10}),
as the case of small slopes.

\section{Discussion and Conclusion}
\label{Conclusion}

In this paper,
we studied the dynamical process of collapsing spheres
in the KMY model,
in which the Schwarzschild radii are time-dependent
due to the back-reaction of Hawking radiation.

First,
we note that,
while an ideal thin shell does not completely evaporate in the KMY model,
a pseudo thin shell does.
The origin of this over-sensitivity on short-distance features
is the higher derivatives needed to determine Hawking radiation.
By properly treating the higher-derivative terms
(in a way analogous to how the Abraham-Lorentz formula
is applied to point charges),
the discrepancy between the ideal thin shell and pseudo thin shell disappears.
All smooth configurations evaporate completely within finite time.

Secondly,
we showed that there are two classes of asymptotic states in the KMY model.
Their initial states are separated by
a critical energy density $\rho_c$ \eqref{rho-c}.
When the initial energy density is much higher than the critical energy density, 
the collapsing shell approaches a thin shell state.
If the initial energy density is much lower than $\rho_c$,
the collapsing shell approaches a slope-1 configuration.

To approach the slope-1 configuration, 
the initial density of the matter has to be smaller than Hawking radiation, 
which is,
for a large range of black-hole masses,
much smaller than that of the current CMB. 
Therefore,
the outer layers of black holes at present
are expected to resemble a (pseudo) thin shell configuration.

Despite the apparent difference between 
the profiles of the slope-1 states and the thin shell states,
both configurations appear to be very similar
from the viewpoint of a distant observer.
The reason is that
the matter at a Planck length under the surface of the shell
is essentially frozen \cite{Ho:2016acf}.
It will be interesting to investigate how these asymptotic states
can be distinguished by certain high-precision observations.

Similarly,
the difference between either of the two asymptotic states
and the conventional model of a black hole with a horizon
is very small to a distant observer
due to the huge redshift factor.
How to distinguish the KMY model and the conventional model
of black holes is a very interesting and important question.
The most salient feature of the KMY model is 
the Planckian scale pressure under the surface of the collapsing matter,
which is at a Planck-scale distance above the incipient horizon
(hence the horizon will not appear).
In contrast,
the near-horizon region of a black hole in the conventional model
is usually assumed to be empty.
This difference is expected to be reflected in the gravitational wave signal
for black-hole mergers.
The gravitational wave signal of black-hole mergers 
has not yet confirmed the existence of the horizon
\cite{Abramowicz:2002vt,Cardoso:2016rao},
while future gravitational wave observations with greater precision
may be able to distinguish the KMY model and the conventional model
\cite{Cardoso:2016oxy,Abedi:2016hgu}.

It should be noted that we have assumed that 
the gravitational force dominates over other forces
in the gravitational collapse
and that the collapsing matter 
is close to the speed of light,
while the pressure of the matter is negligible
(except for that from the quantum effect).
For example, neutron stars have strong Fermi degeneracy pressure 
to resist the gravitational force. 
Thus the black hole formed by the gravitational collapse of a star 
has a minimal mass around a few solar masses.
In this paper, we have assumed that the collapsing matter 
does not have such a large pressure which can stop the collapse. 

\section*{Acknowledgment}

The author would like to thank 
Heng-Yu Chen, Yi-Chun Chin, 
Chong-Sun Chu, Yu-tin Huang, Hsien-chung Kao,
Hikaru Kawai, Ioannis Papadimitriou, Piljin Yi, and Yuki Yokokura,
for discussions.
The work is supported in part by
the Ministry of Science and Technology, R.O.C.
and by National Taiwan University.
The work of Y.M.\ is also supported 
in part by JSPS KAKENHI Grants No.~JP17H06462.

\appendix

\section*{Appendix: Derivative Expansion of Collapsing Shells}
\label{PTS}

Here we verify that,
under certain assumptions about the asymptotic limit,
the slope-1 shells of an arbitrary thickness
(including the pseudo thin shell)
are the only lowest-order solutions
in the derivative expansion
of the formulas of Hawking radiation.
The equations of Hawking radiation in the KMY model are
eqs.\eqref{psi0a}, \eqref{uU-Schwarzian}, \eqref{dR0du}, and \eqref{da0du}.
The lowest-order approximation of eq.\eqref{uU-Schwarzian} is
\footnote{
There is, in fact, an ambiguity in the derivative expansion
depending on which variable 
($\psi_0$ vs. $dU/du$) is used.
}
\be
\{u, U\} \simeq \frac{1}{3}\dot{\psi}_0^2.
\label{uU-Schwarzian-1}
\ee

First of all,
it should be clear that the Hawking radiation 
is dominated by the contributions of 
those layers in the collapsing shell 
which are very close to their Schwarzschild radii.
It is also obvious that,
in terms of the coordinate $u$
suitable for a distant observer,
these layers spend a very long time approaching
their Schwarzschild radii.
The assumption we make is that 
the distribution of these layers
eventually approach a certain universal asymptotic profile
\be
R(a) - a \simeq f(a)
\ee
for a certain given function $f(a)$
that depends only on $a$
and other physical constants.
As the only relevant constant parameter in this model is $\sigma$
(or $\kappa{\cal N}$),
$f(a)$ can always be rewritten as $a g(\sigma/a^2)$
for a certain function $g$.
Since $\sigma/a^2$ is an extremely tiny number,
it is dominated by the leading-order term
in its power expansion in $\sigma/a^2$.
Therefore,
we assume that
\be
R(a) - a \simeq \frac{C}{a^n}
\label{R-a-C}
\ee
at the leading-order in the $(\sigma/a^2)$-expansion
for a constant $C$ and a number $n$.
(This assumption excludes ideal thin shells.)
A generic consequence of this is that
\be
\dot{R}(a) - \dot{a} \simeq 0
\ee
at the leading order.
Taking $u$-derivatives on this equation
and using eq.\eqref{dR0du},
we find
\be
\dot{a}_0 \simeq \dot{R}_0 \simeq - \frac{R_0 - a_0}{2R_0}
\simeq - \frac{C}{2a_0^{n+1}}.
\label{dota0-1}
\ee

According to eq.\eqref{psi0a},
\be
\psi_0 \simeq - \frac{a_0^{n+1}}{(n+1)C}.
\ee
Then eqs.\eqref{da0du} and \eqref{uU-Schwarzian-1} imply
\be
\dot{a}_0 \simeq - \frac{\kappa{\cal N}}{12\pi}\dot{\psi}_0^2
\simeq - \frac{\kappa{\cal N}}{12\pi}\frac{a_0^{2n}}{C^2}\dot{a}_0^2,
\ee
from which we find
\be
\dot{a}_0 \simeq - \frac{12\pi}{\kappa{\cal N}}\frac{C^2}{a_0^{2n}}.
\label{dota0-2}
\ee

The agreement between eqs.\eqref{dota0-1} and \eqref{dota0-2}
demands that
\begin{align}
n &= 1,
\\
C &= \frac{\kappa{\cal N}}{24\pi}.
\end{align}
Therefore the asymptotic profile \eqref{R-a-C}
is precisely the slope-1 configuration
(without restriction on its thickness
so pseudo thin shells are included).
For the self-consistency of this result, 
one can check that
the higher-order terms ignored in eq.\eqref{uU-Schwarzian-1}
are indeed much smaller than the lower-order terms.


\end{document}